\documentclass[aps,prl,twocolumn]{revtex4-1}

\usepackage{amsmath,amssymb,bm,graphicx,epsfig,psfrag,ulem,color,natbib}

\newcommand{\bs}{\mathbf {s}}
\newcommand{\br}{\mathbf {r}}
\newcommand{\bv}{\mathbf {v}}

\newcommand{\bom}{{\mbox{\boldmath $\omega$}}}
\begin{document}

\title{Quantum turbulent velocity statistics and quasiclassical limit}

\author{A.W. Baggaley and C.F. Barenghi}
\affiliation{School of Mathematics and Statistics, 
Newcastle University, Newcastle upon Tyne, NE1 7RU, United Kingdom}

%\date{\today}

\begin{abstract}
Two research groups have measured turbulent velocity statistics
in superfluid helium using different techniques. The results
were in conflict: one experiment revealed 
Gaussian distributions (as observed in ordinary
turbulence), the other experiment determined power-laws.
To solve the apparent puzzle, we numerically model quantum turbulence
as a tangle of vortex filaments,
and conclude that there is no contradiction between the 
two experiments. The transition from Gaussian to
power-law arises from the different length scales which are
probed using the two techniques. 
We find that the average distance between the
quantum vortices marks the separation between quantum and quasi-classical
length scales.
\end{abstract}

%Something or other
\pacs{67.25.dk Vortices and turbulence,\\
47.32.C Vortex dynamics (fluid flow)}

\maketitle

Turbulence in superfluid helium ($^4$He)~\cite{Manchester,Prague}
consists of a tangle of discrete, thin vortex
filaments, each carrying one quantum of 
circulation $\kappa=h/m$ (where $h$ is Planck's constant and $m$ the
mass of one atom). 
This `quantum turbulence'~\cite{Vinen,Halperin} is also studied in superfluid 
$^3$He~\cite{Lancaster,Helsinki}
and in atomic Bose-Einstein condensates~\cite{Bagnato}. 
The absence of viscosity and the
discrete nature of the vorticity distinguish quantum turbulence from
ordinary (classical) turbulence. Nevertheless,
experiments have also shown remarkable similarities:
the same pressure drop along and pipes~\cite{VanSciver-pressure}, 
drag on a moving body~\cite{VanSciver-sphere}, 
vorticity decay~\cite{Oregon}, and  
distribution of kinetic energy over the length 
scales (Kolmogorov energy spectrum)~\cite{Tabeling,Salort2010}.
The relation between quantum turbulence and classical turbulence 
has thus become the focus of attention.

This report is concerned with the statistical properties of
the components of the turbulent superfluid velocity field $\bv$.
In ordinary turbulence, experiments~\cite{Noullez} and
numerical simulations~\cite{Vincent,Gotoh} confirm that
velocity statistics are
Gaussian.
The usual interpretation of this property is the 
following~\cite{Davidson}. At any time $t$, 
the velocity field $\bv$ at the point $\br$
is determined by the vorticity field $\bom=\nabla \times \bv$ via the 
Biot-Savart law \cite{Saffman}
\begin{equation}
\bv(\br,t)=\frac{1}{4 \pi} \int 
\frac{\bom(\br',t) \times (\br -\br')}{\vert \br -\br'\vert^3}\, d\br'.
\label{eq:BSomega}
\end{equation}

\noindent
Thus, if the point $\br$ is surrounded by many, randomly oriented vortical
structures, Gaussianity results from the application of the
Central Limit Theorem.

In the last few years, the development of particle tracking
visualization techniques suitable
for liquid helium~\cite{VanSciver,Bewley}
has opened the way to the measurement of quantum turbulence 
statistics.  Using solid hydrogen tracers,
Paoletti et al.~\cite{Paoletti} at the University of Maryland deduced that, in
superfluid helium, the 
velocity components of the turbulent velocity follow power-law
distributions.  Large velocity fluctuations are thus relatively more
frequent than in ordinary turbulence, where these distributions are
Gaussian~\cite{note}.
The authors suggested that power-laws distributions arise
from vortex reconnections, clearly high-velocity events.

The experiment of Paoletti et al. motivated theoretical work on the
problem. White et al.~\cite{White} solved numerically the
Gross-Pitaevskii equation for a Bose-Einstein condensate and
determined velocity distributions in a variety of
quantum vortex systems at zero temperature: 
two-dimensional and three-dimensional
trapped atomic condensates, three-dimensional homogeneous condensates,
Onsager's two-dimensional vortex gas. 
In all systems, they found power-law distributions,
in agreement with Paoletti et al.~\cite{Paoletti}, even
in the absence of vortex reconnections.  Using results from previous work 
by Min and Leonard~\cite{Leonard} on the numerical analysis of vortex
methods (later confirmed by Weiss et al.~\cite{Weiss}),
White et al.~\cite{White} 
concluded that the non-Gaussianity arises from the singular
nature of the quantum vorticity, which causes an anomalously slow
convergence to Gaussian behaviour. 

Further theoretical work at non-zero temperatures
seemed to strengthen this conclusion.
Baggaley and Barenghi~\cite{Baggaley-vortexdensity}
used the vortex filament model to numerically study
intense vortex tangles. They confirmed the
Kolmogorov spectrum and the power-laws distributions of the
velocity components.
The vortex filament model was also used by Adachi and
Tsubota~\cite{Adachi}, who found power-law distributions
in counterflow turbulence (quantum turbulence driven by a heat flow).

In a recent experiment, Salort et al.~\cite{Salort2011} at CNRS Grenoble
measured the turbulent velocity in a superfluid wind tunnel using a 
small Pitot tube. They confirmed the Kolmogorov spectrum, but 
reported Gaussian velocity distributions, in apparent
contradiction with Paoletti et al.~\cite{Paoletti}
and with the numerical simulations~\cite{White,Baggaley-vortexdensity,Adachi}.

To solve this puzzle, firstly we consider the size of the
probe relative to the distance between vortices in the two experiments.
The radius $a$ of the tracer particles 
used by Paoletti et al. was  $a\approx 10^{-4}~\rm cm$. 
The turbulence was created by a heat flux $\dot Q$ which
drove a counterflow current~\cite{Adachi2010} 
$v_{ns}=\vert v_n-v_s \vert={\dot Q}/(\rho_s S T)$ between superfluid
and normal fluid velocity components $v_n$ and $v_s$, where
$\rho_s$ is the superfluid density and $S$ the specific entropy. 
For example, at $T=1.9~\rm K$ the largest heat flux 
was ${\dot Q}=0.17~\rm W/cm^2$,
corresponding to $v_{ns}=1.5~\rm cm/s$, with
$\rho_s=0.0831~\rm g/cm^3$ and $S=0.709~\rm J/(gK)$. Using the relation
$L = \gamma^2 v_{ns}^2$ 
with $\gamma \approx 140~\rm s/cm^2$
as reported in Ref.~\cite{Adachi2010}, one finds 
$\ell \approx 4.7 \times 10^{-3}~\rm cm$, where $L$ is the vortex line density (vortex length per unit volume)
and $\ell\approx L^{-1/2}$. Paoletti et al.~\cite{Paoletti}
took their velocity measurements during the decay of the turbulence, 
after switching off the heater, so the actual inter-vortex distance 
must have been larger than the above value of $\ell$. 
We conclude that, in the experiment of Paoletti et al.~\cite{Paoletti},
$a/\ell \ll 1$;
this means that, although the tracer particles must have altered
in some measure the motion of the vortex lines, the 
velocity field was probed at very small length scales, between $\ell$ and $a$.

In the experiment of Salort et al.~\cite{Salort2011},
the nozzle of the Pitot tube 
had diameter $0.06~\rm cm$. The authors
estimated~\cite{Roche} that the inter-vortex separation was 
between $\ell \approx 5 \times 10^{-4}$ and $20 \times 10^{-4} \rm cm$.
We conclude that in this experiment
$a/\ell \gg 1$ (even without considering that the
effective resolution in the streamwise direction, set by time dynamics
of the probe, was perhaps much larger than $a$~\cite{Roche}).
In summary, the experiment of Paoletti et al.~\cite{Paoletti}
probed the velocity field between vortices, at length
scales less than $\ell$, whereas the experiment
of Salort et al.~\cite{Salort2011} 
detected motion at scales larger than $\ell$, containing many vortex
lines.

Secondly, we perform the following numerical
calculation. In both experiments
the length scales involved were much larger than the vortex core
radius, $a_0=10^{-8}~\rm cm$, so it is appropriate
to use the vortex filament model~\cite{Schwarz}.
We assume that the vorticity is concentrated in space curves
$\bs(\xi,t)$ around which the circulation is $\kappa=10^{-3}\rm cm^2/s$.
Eq.~(\ref{eq:BSomega}) reduces to

\begin{equation}
\frac{d{\bf s}}{dt}=-\frac{\kappa}{4 \pi} \oint_{\cal L} \frac{(\bs-\br) }
{\vert \bs - \br \vert^3}
\times {\bf d}\br,
\label{eq:BS}
\end{equation}

\noindent
where the line integral extends over all vortex lines. The numerical
techniques to discretize the vortex lines, to compute the time
evolution, to de-singularize the Biot-Savart integral and to 
algorithmically perform reconnections when two vortex lines become
close to each other are described 
in the literature~\cite{Schwarz,Adachi} or in our
 previous papers~\cite{Baggaley-cascade,Baggaley-vortexdensity}.
The computational domain is a periodic box of size $D=0.075~\rm cm$. 
The distance between discretization points along the vortex
filaments is held at approximately $\delta/2$ where $\delta=0.001~\rm cm$.
To speed-up the evaluation of Biot-Savart integrals 
we use a tree method~\cite{Baggaley-long}
with critical opening angle $\theta=0.4$.
For the sake of simplicity, we assume that the temperature is $T=0$.
This means that no external forcing is needed to sustain the turbulence,
because the total kinetic energy is 
conserved during the evolution (at least for the time scale of
interest here; the finite discretization and the reconnection
algorithm introduce small energy losses which model phonon emission
as described in Ref.~\cite{Baggaley-cascade}).

\begin{figure}
\begin{center}
\includegraphics[width=0.45\textwidth]{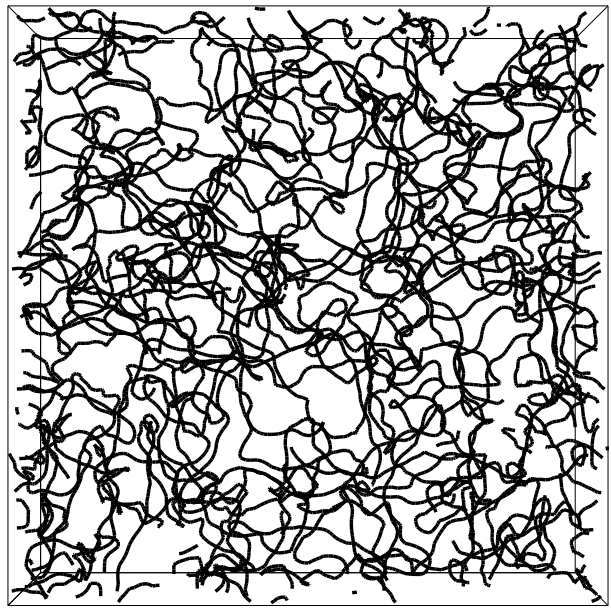}     
\caption{
Snapshot of the turbulent vortex tangle at $t=0.1\,$s after the vortex line
density has saturated, the box size $D=0.075~\rm cm$. 
}
\label{fig1}
\end{center}
\end{figure}

The initial condition at time $t=0$ consists of 100 straight, randomly oriented, vortex 
filaments discretized over $N=10^4$ Lagrangian points $\bs_j$
($j=1,\cdots N$).
During the evolution the vortex lines interact, become curved,
reconnect, and a vortex tangle is quickly formed. We find that  
the vortex line density $L$
initially grows, then saturates to the value
$L\approx 2.04 \times 10^4~\rm cm^{-2}$.
We check that the average curvature saturates too.  
Fig.~\ref{fig1} shows the vortex
tangle when we stop the calculation at $t=0.1~\rm s$. 
We examine
the energy spectrum $E(k)$, defined by
\begin{equation}
E=\frac{1}{V} \int \frac{1}{2} \vert \bv \vert^2 dV=\int_0^{\infty} E(k)\,dk,
\label{eq:spectrum}
\end{equation}

%%%%%%%%%%%%%%%%%%%%%%%%%%%%%%%%%%%%%%%%%%%%%%%%%%%%%%%%%%%%%%%%%%%%%%%%%
\begin{figure}
\begin{center}
\includegraphics[width=0.45\textwidth]{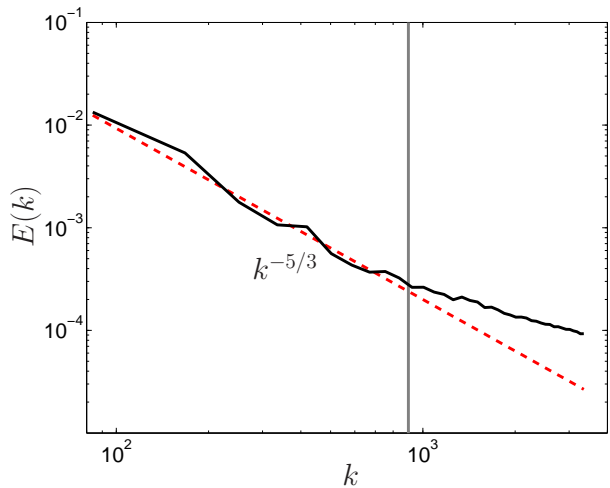}     
\caption{
(Color online) Energy spectrum $E_k$ (arbitrary units) vs wavenumber $k$ (${\rm cm}^{-1}$) 
corresponding to Fig.~\ref{fig1}. The dashed line is the Kolmogorov
scaling $E_k \sim k^{-5/3}$; the solid grey line shows wavenumber corresponding to the  intervortex spacing $k_{\ell}\approx 900~\rm cm^{-1}$.
}
\label{fig2}
\end{center}
\end{figure}

\begin{figure}
\begin{center}
\includegraphics[width=0.4\textwidth]{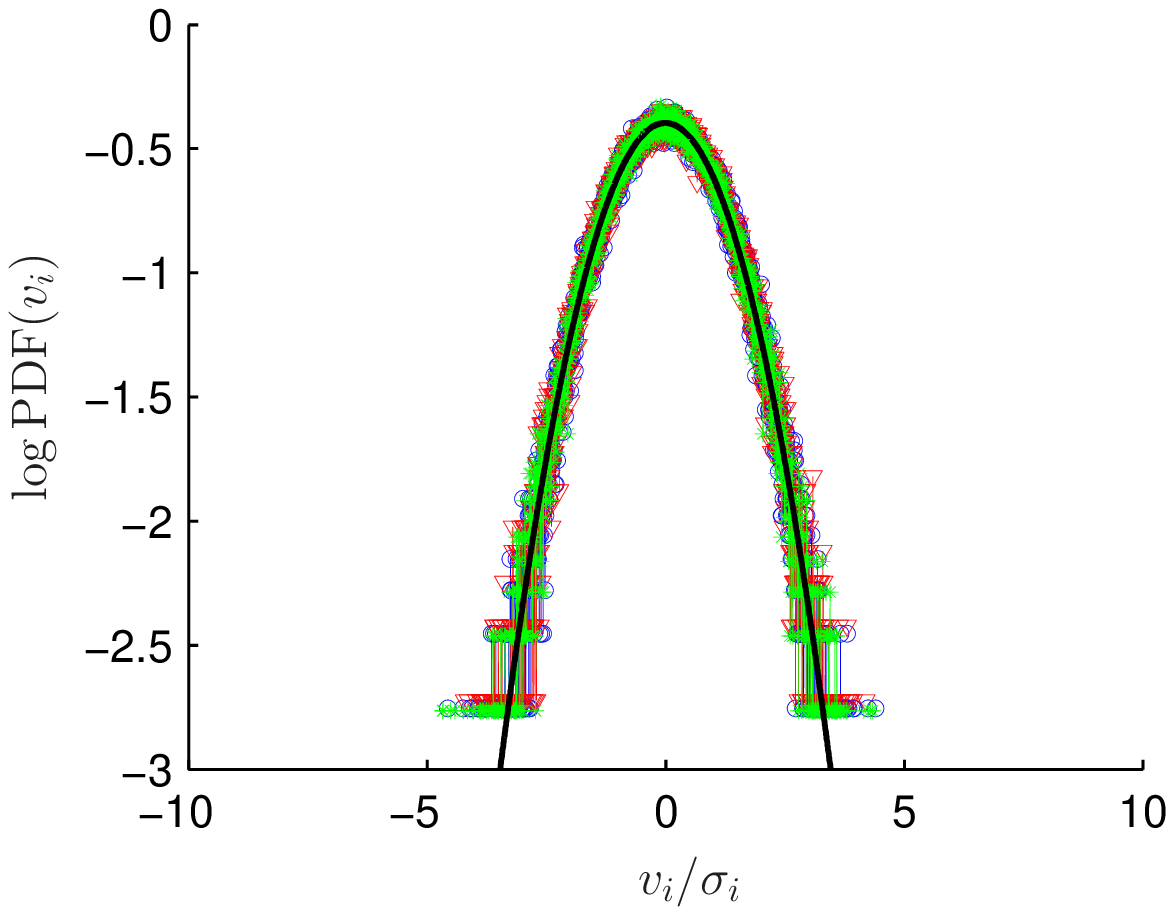}     
\includegraphics[width=0.4\textwidth]{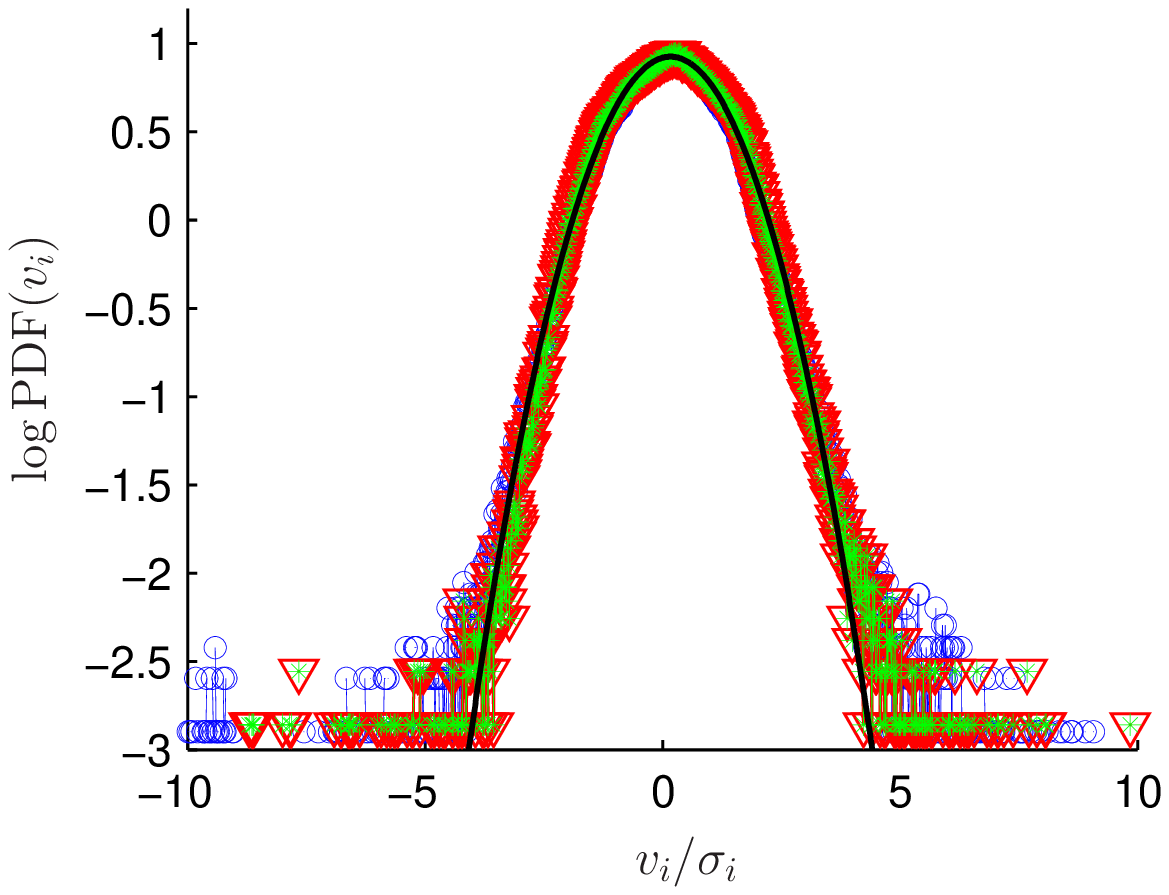}     
\includegraphics[width=0.4\textwidth]{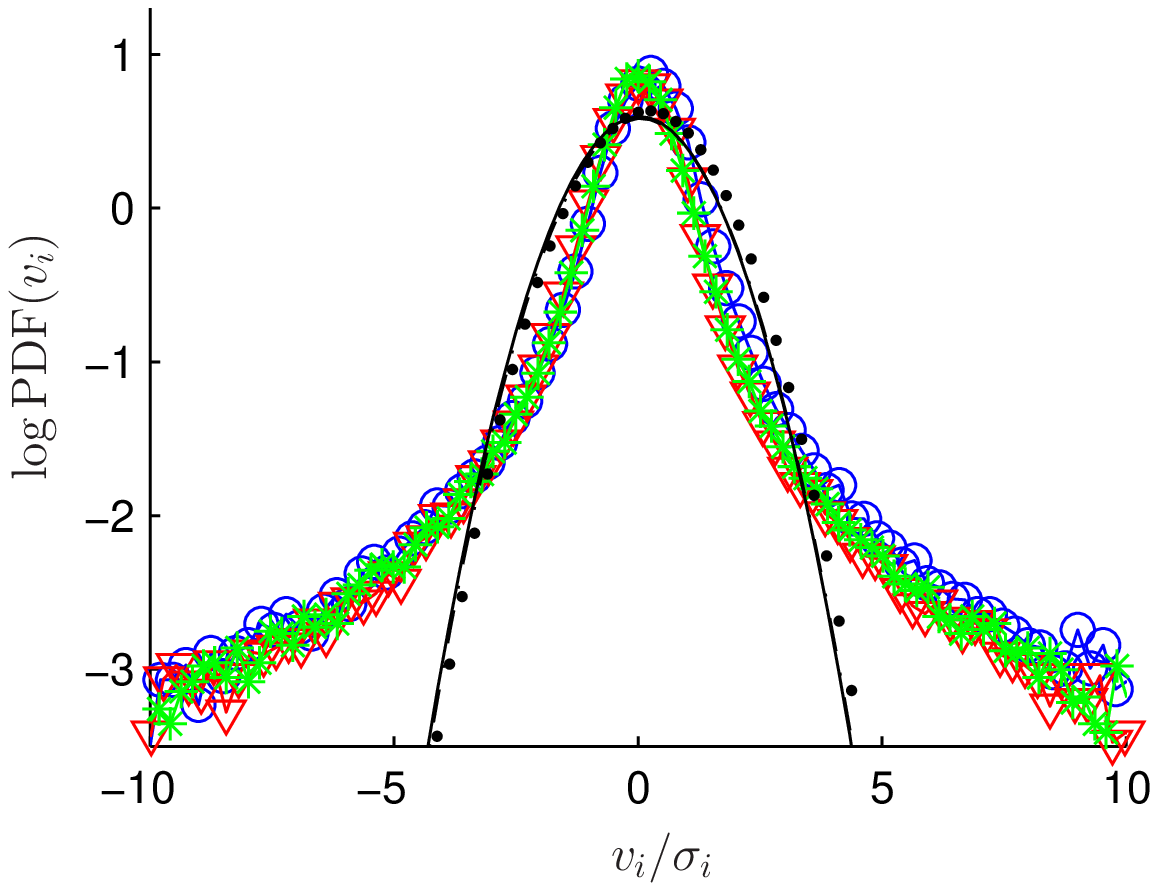}     
\caption{
(Color online) Probability density functions (PDF) of turbulent
velocity components $v_i$ ($i=x,y,z)$ vs $v_i/\sigma_i$
calculated by averaging over regions of size $\Delta=2 \ell$ (top),
$\Delta=\ell$ (middle) and $\Delta=\ell/6$ (bottom). (Green) asterisks,
(blue) circles and (red) triangles refer respectively to $i=x$, 
$i=y$ and $i=z$ components. 
The solid lines ares the Gaussian fits, defined by Eq.~\ref{eq:Gaussian},
where $\bar{\sigma}=0.041$ (top), $0.048$ (middle) and 
$0.082~\rm cm/s$ (bottom) respectively. 
}
\label{fig3}
\end{center}
\end{figure}

%%%%%%%%%%%%%%%%%%%%%%%%%%%%%%%%%%%%%%%%%%%%%%%%%%%%%%%%%%%%%%%%%%%%%%%%%

\noindent
where $E$ is the total energy per unit mass. Fig.~\ref{fig2} shows that most of the energy is contained in the large
scales, and that the spectrum is qualitatively consistent with
the Kolmogorov scaling $E_k \sim k^{-5/3}$ in the region
$k \ll 2 \pi/\ell \approx 900~\rm cm^{-2}$, in agreement with 
experiments \cite{Tabeling,Salort2010} and with
previous numerical calculations 
\cite{Nore,Tsubota-vd,Baggaley-vortexdensity,Tsubota-GP}. The shallow
spectrum at large values of $k$ is also consistent with existing
similar calculations~\cite{Tsubota-vd}.
Secondly, we examine the
velocity components and compute their probability density functions
(normalized histograms, or PDF for short). 
Rather than sampling the values at points
of the domain, we compute averages over regions of size $\Delta$,
varying $\Delta$. 
This is done by calculating the velocity field, induced by the vortices, on
a $128^3$ Cartesian mesh, and then averaging cells over the appropriate length scale.
The result is shown in Fig.~\ref{fig3}, where
we compare the data against their Gaussian fits
\begin{equation}
{\rm gPDF}(v_i)=\frac{1}{\sqrt{2 \pi \sigma^2}}
{\rm exp}(-(v_i-\mu)^2/(2 \sigma^2)),
\label{eq:Gaussian}
\end{equation}
%%%%%%%%%%%%%%%%%%%%%%%%%%%%%%%%%%%%%%%%%%%%%%%%%%%%%%%%%%%%%%%%%%%%%%%%%
%%%%%%%%%%%%%%%%%%%%%%%%%%%%%%%%%%%%%%%%%%%%%%%%%%%%%%%%%%%%%%%%%%%%%%%%%

\noindent
where $\mu \approx 0$ and $\sigma$ is the standard deviation.
It is apparent that,
if $\Delta <\ell$, the PDFs are not Gaussian; by fitting
${\rm PDF}(v_i) \sim v_i^{-b}$ we obtain $b=3.2$,
$3.2$ and $3.1$ for $i=x,y,z$ respectively, in agreement with
the power-law statistics found
experimentally by Paoletti et al.~\cite{Paoletti} and numerically
by White et al~\cite{White}, 
Baggaley and Barenghi~\cite{Baggaley-vortexdensity}, 
and Adachi and Tsubota~\cite{Adachi}. 
If $\Delta >\ell$, however, the distributions are Gaussian,
and we recover the same (classical)
statistics measured by Salort et al.~\cite{Salort2011}.

In conclusion, the different velocity statistics observed by the
Maryland and Grenoble experiments are consistent with each other.
Taken together, these experiment and this work support 
the interpretation that, at scales larger than
$\ell$, quantum turbulence exhibits quasi-classical behaviour
(in terms of both energy spectrum and velocity statistics),
whereas at scales smaller
than $\ell$ (but still orders of magnitude larger than the quantum
coherence length $a_0$), the discrete nature of quantized vorticity affects
the distribution of energy over the length scales and the frequency of
high-velocity events.

%%%%%%%%%%%%%%%%%%%%%%%%%%%%%%%%%%%%%%%%%%%%%%t%%%%%%%%%%%%%%%%%%%%%%%%%%
%%% FIG 3

\begin{acknowledgments}
We thank J. Salort and P.E. Roche for providing us with experimental data
and comments in advance of publication.
This work was funded by the HPC-EUROPA2 project
228398, with the support of the European Community
(Research Infrastructure Action of the FP7), and  by the
Leverhulme Trust (Grants F/00125/AH and RPG-097).
\end{acknowledgments}

\end{document}